\documentclass[twocolumn]{aastex631} 

\usepackage{amsmath}
\usepackage{color}

\def\gsim{\mathrel{\rlap{\lower 4pt \hbox{\hskip 1pt $\sim$}}\raise 1pt
\hbox {$>$}}}
\def\lsim{\mathrel{\rlap{\lower 4pt \hbox{\hskip 1pt $\sim$}}\raise 1pt
\hbox {$<$}}}

\begin{document}

\title{
Formation of circumstellar material during double-white-dwarf mergers and the early excess emissions in Type Ia supernovae}

\correspondingauthor{Yusuke Inoue}
\email{yusuke@kusastro.kyoto-u.ac.jp}

\author[0009-0004-3148-0462]{Yusuke Inoue}
\author[0000-0003-2611-7269]{Keiichi Maeda}
\affiliation{Department of Astronomy, Kyoto University, Kitashirakawa-Oiwake-cho, Sakyo-ku, Kyoto, 606-8502. Japan}

\author[0000-0002-3933-7861]{Takashi Nagao}
\affiliation{Department of Physics and Astronomy, University of Turku, FI-20014 Turku, Finland}
\affiliation{Aalto University Mets\"ahovi Radio Observatory, Mets\"ahovintie 114, 02540 Kylm\"al\"a, Finland}
\affiliation{Aalto University Department of Electronics and Nanoengineering, P.O. BOX 15500, FI-00076 AALTO, Finland}
\affiliation{National Astronomical Observatory of Japan, National Institutes of Natural Sciences, 2-21-1 Osawa, Mitaka, Tokyo 181-8588, Japan}

\author[0000-0002-9350-6793]{Tatsuya Matsumoto}
\affiliation{Department of Astronomy, School of Science, The University of Tokyo, Bunkyo-ku, Tokyo 113-0033, Japan}

\begin{abstract}
Early excess emission observed in Type Ia supernovae (SNe Ia) within $\sim1$ day of explosion provides a critical window into their progenitor systems.
In the present study, we investigate formation of the circumstellar matter (CSM) in double white-dwarf (WD) mergers. We further study the interaction between the CSM and the SN ejecta. 
We first model the orbital evolution and super-Eddington mass transfer/ejection in the double WD systems.
We then conduct hydrodynamical and light-curve (LC) simulations of the SN-CSM interaction, assuming a prompt SN Ia explosion in a context of the carbon-ignited violent merger (C-ignited VM).
Our simulations show that at the moment of the merger, the binary system has the CSM distribution following $\rho_{\mathrm{CSM}}\simeq D(r/10^{14}\ \mathrm{cm})^{-3.5}\ (D\simeq 10^{-14}\text{--}10^{-13}\ \rm g\ cm^{-3})$.
The simulated LCs reproduce the early flux excesses across optical to UV bands, as well as their color evolution, observed in the VM candidates, i.e., 03fg/02es-like SNe Ia. 
This supports that 03fg/02es-like objects originate from the VM explosions.
We also discuss the case of the helium-ignited VM, which might be realized in some WD-WD mergers depending on the He content in the system. Focused here is the timing when the explosion is initiated, and we find that the explosion is initiated after the companion WD is, at least partially, tidally disrupted also in this case; we thus expect the formation of the CSM through the mass transfer phase also for the helium-ignited VM scenario. 
\end{abstract}


\section{Introduction} \label{sec:intro}
    Type Ia supernovae (SNe Ia) serve as a cosmic standardized candle owing to their remarkably uniform light curves \citep[][]{1977SvA....21..675P,1993ApJ...413L.105P}.
    However, a growing number of peculiar SNe Ia challenge this standard picture \citep[e.g., ][for reviews]{maeda2016IJMPD..2530024M,2025A&ARv..33....1R}, suggesting a diversity in their progenitor channels.
    Indeed, despite extensive observational and theoretical efforts, the progenitor channels for SNe Ia are still unclear. 
    The double-degenerate (DD) system, which consists of two white dwarfs (WDs), is one of the promising progenitor candidates \citep[e.g.,][]{1984ApJS...54..335I,1984ApJ...277..355W}.\par
    Several previous studies have conducted hydrodynamics simulations of the double C/O WDs' merger, and reported the possibility of the prompt and direct detonation of carbon, called the carbon-ignited violent merger (hereafter VM), as one of the final products of the DD systems \citep[e.g.,][]{2010Natur.463...61P,2011A&A...528A.117P,2012ApJ...747L..10P,2015ApJ...807..105S,2016ApJ...821...67S,2025arXiv251011781P}.
    Furthermore, radiative transfer simulations have revealed that the VM systems can reproduce properties of a few subclasses of SNe Ia, depending on the WD masses and the viewing angle \citep[e.g.,][]{2010Natur.463...61P,2012ApJ...747L..10P,2014ApJ...785..105M}. \par

    \begin{table*}[t]
    \begin{center}
    \caption{
    03fg- and 02es-like objects accompanied by the early flux excess.
    The following objects are also associated with possible early-flux excesses but not included here, given the small numbers of data points or large statistical errors in their early-phase observations; ASASSN-15pz \citep[03fg-like, ][]{2019ApJ...880...35C} and LSQ 12gpw \citep[03fg-like, ][]{2018ApJ...865..149J}, iPTF 14dpk \citep[02es-like, ][]{2016ApJ...832...86C,2018ApJ...865..149J} and ASASSN-20jq/SN 2020qxp \citep[02es-like, ][]{2025arXiv250104086B}. 
    \label{table:obsdata}}
    {
    \begin{tabular}{cccc} \hline
       Object 
       & Type 
       & Instruments (band)  
       & References \\ 
       \hline
            SN 2020hvf & 03fg-like & Tomo-e (clear)   & \citet{2021ApJ...923L...8J}   \\
            SN 2021zny & 03fg-like & ZTF (g)          & \citet{2023MNRAS.521.1162D}   \\
            SN 2022ilv & 03fg-like & ATLAS (o)        & \citet{2023ApJ...943L..20S}   \\
            SN 2022pul & 03fg-like & ASAS-SN (g)      & \citet{2024ApJ...960...88S}   \\ 
            iPTF 14atg & 02es-like & PTF (r)          & \citet{2015Natur.521..328C}   \\
            SN 2016jhr & 02es-like & HSC (g)          & \citet{2017Natur.550...80J}   \\
            SN 2019yvq & 02es-like & ZTF (g), LCO (g) & \citet{2020ApJ...898...56M,2021ApJ...919..142B}  \\
            SN 2022vqz & 02es-like & ZTF (g), TNT (g)          & \citet{2024MNRAS.527.9957X}   \\
            SN 2022ywc & 02es-like & ATLAS (o)        & \citet{2023ApJ...956L..34S}   \\
       \hline
    \end{tabular}
    }
    \end{center}
    \end{table*}
    
    Observationally, the over-luminous SNe Ia, called SN 2003fg-like (03fg-like) objects, have been proposed as possible DD and VM candidates, based on their high peak luminosity, presence of absorption lines associates with unburnt carbon, a narrow [OI] emission line 
    in the nebula phase, and the high polarization ($\sim1-2\ \%$) \citep[e.g,][]{2021ApJ...923L...8J,2022ApJ...927...78D,2023MNRAS.521.1162D,2023ApJ...943L..20S,2024ApJ...966..135K,2024ApJ...960...88S,2024A&A...687L..19N}.
    An interesting feature shared by 03fg-like objects is the flux excess in their LCs within a few days after the explosion (hereafter the early flux excess) (Table \ref{table:obsdata}).
    \par 
    Some sub-luminous SNe Ia, called SN 2002es-like (02es-like) objects, have also been proposed to arise from the VM based on their slowly evolving LCs and spectra, a narrow [OI] emission line in the nebular spectra, and the carbon absorption lines \citep[][]{2011MNRAS.418..747M,2012ApJ...751..142G,2013ApJ...770L...8P,2013ApJ...775L..43T,2013ApJ...778L..18K,2016MNRAS.459.4428K,2023ApJ...950...17L,2025arXiv251011781P}.
    Like the SN 2003fg-like objects, the early flux excess is also a common feature in 02es-like objects (Table \ref{table:obsdata}).\par
    For the origin of the early flux excess, several mechanisms have been proposed; the SN-CSM interaction, the companion interaction, the unusual $\mathrm{^{56}Ni}$ distribution, and the He detonation scenarios \citep[e.g.,][]{2010ApJ...708.1025K,2016ApJ...826...96P,2017MNRAS.472.2787N,2017Natur.550...80J,2018ApJ...861...78M}.
    It might indeed be a mixture of these mechanisms, but no robust conclusion has been reached as for which class of SNe Ia with the early flux excess is associated with which mechanism.
    In the present work, we focus on the scenario where an additional energy input is provided by the interaction between the SN ejecta and CSM, which has been suggested to be a promising scenario for the early flux excesses seen in 03fg- and 02es-like objects \citep[e.g.,][]{2021ApJ...923L...8J,2023MNRAS.521.1897M,2023ApJ...956L..34S,2024ApJ...966..139H}.\par
    
    The formation of the CSM in a double WD merger system, including the VM channel, is still under discussion. Various scenarios proposed so far assume that the CSM is originated in the tidally disrupted debris of the less massive (donor) WD component; the tidal tail ejection \citep[][]{2013ApJ...772....1R,2014MNRAS.438...14D,2015ApJ...807...40T,2025arXiv251011781P}, post-merger viscous evolution of the disk formed by the completely tidally disrupted WD \citep[][]{2012ApJ...747L..10P,2012ApJ...748...35S,2012MNRAS.427..190S}, and the wind from the disk \citep[][]{2015MNRAS.447.2803L,2017MNRAS.470.2510L}. These scenarios have a common problem; these processes are expected to take place within the dynamical timescale of the DD system ($\sim10\text{--}100$ s) before the carbon detonation \citep[e.g.,][]{2010Natur.463...61P,2011A&A...528A.117P,2010ApJ...709L..64G}, which is too short for the ejected materials to reach an extended scale as a CSM to explain the early flux excess. Thus, the unreasonably long lag time of $>10^{4}\text{--}10^{5}$ s between the onset of disruption of the donor WD and the SNe Ia explosion must be hypothesized. 
    In the present study, we suggest a new scenario for the creation of the CSM in 03fg- and 02es-like objects in the context of the VM scenario, where the CSM is associated with a wind driven by the rapid mass transfer before the tidal disruption of the donor WD.\par
    This paper is structured as follows. 
    The DD binary evolution model is described in Section \ref{sec:method}. 
    The results of the orbital evolution and the expected CSM structures are presented in Section \ref{sec:Results}. 
    In Section \ref{sec:vmLC}, we show the synthetic optical and UV LCs arising from the interaction between the SN ejecta and the CSM, where the comparison with observational data is also conducted.
    In Section \ref{sec:dis}, further implications from our model, as well as some limitations and caveats, are discussed.
    In Section \ref{sec:summary}, we summarize our findings.
    In addition, the notations of physical constants are shown in Section \ref{sec:phyconst}. \par

\section{Models of binary evolution and CSM}\label{sec:method}
    In this section, we describe our binary model including the mass transfer and the wind, for double WD binary systems in the super-Eddington mass transfer phase.
    In the high mass-ratio DD binary system, dynamically unstable mass transfer and subsequent DD merger are expected \citep[e.g.,][]{1999A&A...349L..17H,2004MNRAS.350..113M}.
    \subsection{Binary evolution with the wind mass loss} \label{subsec:separation}
        The total orbital angular momentum ($J_{\mathrm{orb}}$) is expressed as follows:
        \begin{equation}\label{eq:totalangularmomentum}
        J_{\mathrm{orb}}=M_{\mathrm{A}}M_{\mathrm{D}}\left(\frac{Ga}{M_{\mathrm{A}}+M_{\mathrm{D}}}\right)^{1/2},
        \end{equation}
        where $M_{\mathrm{A}}$ and $M_{\mathrm{D}}$ are the masses of the accreting and donor WDs, and $a$ is the orbital separation. By logarithmic differentiation one finds
        \begin{equation}\label{eq:orbitalevolution}
            \frac{\dot{a}}{a}
            =-2\frac{\dot{M}_{\mathrm{A}}}{M_{\mathrm{A}}} -2 \frac{\dot{M}_{\mathrm{D}}}{M_{\mathrm{D}}} + \frac{\dot{M}_{\mathrm{A}}+\dot{M}_{\mathrm{D}}}{M_{\mathrm{A}}+M_{\mathrm{D}}}+2\frac{\dot{J}_{\mathrm{orb}}}{J_{\mathrm{orb}}}, 
        \end{equation}
        where $\dot{M}_{\mathrm{D}}$ expresses the mass transfer rate from the donor WD to the accreting WD.\par
        For $\dot{M}_{\mathrm{D}}$, we employ an adiabatic model that is valid for a rapid mass transfer expected in the current context \citep{2004MNRAS.350..113M}. 
        Within this model, the mass transfer rate is given by the donor radius ($R_{\mathrm{D}}$) and the Roche lobe radius around the donor  ($R_{\mathrm{L}}$) as follows:
        \begin{equation}\label{eq:masstransrate}
            \begin{split}
                \dot{M}_{\mathrm{D}}
                =&\frac{8\pi^{3}}{9}
                \left(\frac{5Gm_{\mathrm{e}}}{h^{2}}\right)^{3/2}
                (2m_{\mathrm{n}})^{5/2}
                \left(\frac{4\pi a^{3}}{G(M_{\mathrm{A}}+M_{\mathrm{D}})}\right)^{-1/2}\\
                &\times
                \left( \frac{3 \mu M_{\mathrm{D}}}{5r_{\mathrm{L}}R_{\mathrm{D}}} \right)^{3/2}
                (a_{\mathrm{D}}(a_{\mathrm{D}}-1))^{-1/2}\\
                &\times
                (R_{\mathrm{D}}-R_{\mathrm{L}})^{3},
            \end{split}
        \end{equation}
        where $\mu=M_{\mathrm{D}}/(M_{\mathrm{A}}+M_{\mathrm{D}})$, $r_{\mathrm{L}}=R_{\mathrm{L}}/a$, and $a_{\mathrm{D}}=\mu/x_{\mathrm{L1}}^{3}+(1-\mu)/(1-x_{\mathrm{L1}})^{3}$ \citep[e.g.,][]{2004MNRAS.350..113M}.
        Here $x_{\mathrm{L1}}=(0.696q^{1/3}-0.189q^{2/3})/(1+0.014q)$ is an approximated distance between the donor WD and the inner Lagrangian point in units of $a$ \citep[e.g.,][]{1999A&A...349L..17H} with the mass ratio of $q=M_{\mathrm{D}}/M_{\mathrm{A}}$.
        For $R_{\mathrm{L}}$, we employ the approximation from \citet{1983ApJ...268..368E} as follows: 
        \begin{equation}\label{eq:eggleton}
            \frac{R_{\mathrm{L}}}{a}=\frac{0.49q^{2/3}}{0.6q^{2/3}+\mathrm{ln}(1+q^{1/3})}.
        \end{equation}
        For $R_{\mathrm{D}}$, we use the Eggleton's zero-temperature mass ($M$)-radius ($R$) relation \citep{1988ApJ...332..193V,2004MNRAS.350..113M}:
        \begin{equation}\label{eq:massradiusrela}
            \begin{split}
                R=
                &0.0114R_{\odot}\left[ \left(\frac{M}{M_{\mathrm{Ch}}}\right)^{-2/3} - \left(\frac{M}{M_{\mathrm{Ch}}}\right)^{2/3} \right]^{1/2}\\
                &\times
                \left[1+ 3.5\left(\frac{M}{M_{\mathrm{p}}}\right)^{-2/3} + \left(\frac{M}{M_{\mathrm{p}}}\right)^{-1} \right]^{-2/3},
            \end{split}
        \end{equation}
        where $M_{\mathrm{Ch}}=1.44M_{\odot}$ and $M_{\mathrm{p}}=5.7\times 10^{-4}M_{\odot}$.
        Note that this relation holds only for a WD under the hydrostatic equilibrium.
        We confirm that this always applies in the present situation, since the dynamical timescale ($\sim\sqrt{R_{\rm D}^3/GM_{\rm D}}$) is shorter than the timescale for the mass loss ($\sim M_{\rm D}/\dot{M}_{\rm D}$).
        \par
        
        \begin{figure*}[t]
        \begin{minipage}[b]{0.1\linewidth}
        \centering
        \includegraphics[keepaspectratio, scale=0.39]{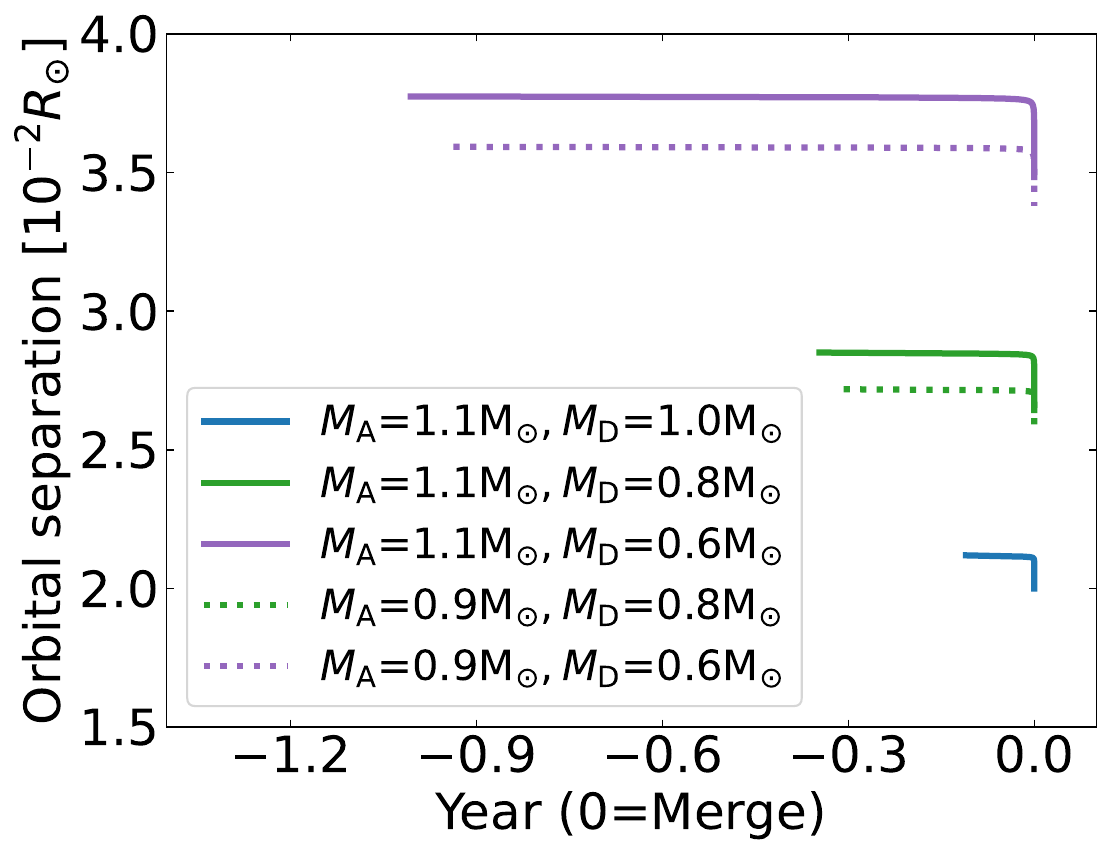}
        \end{minipage}
        \hfill
        \begin{minipage}[b]{0.5\linewidth}
        \centering
        \includegraphics[keepaspectratio, scale=0.39]{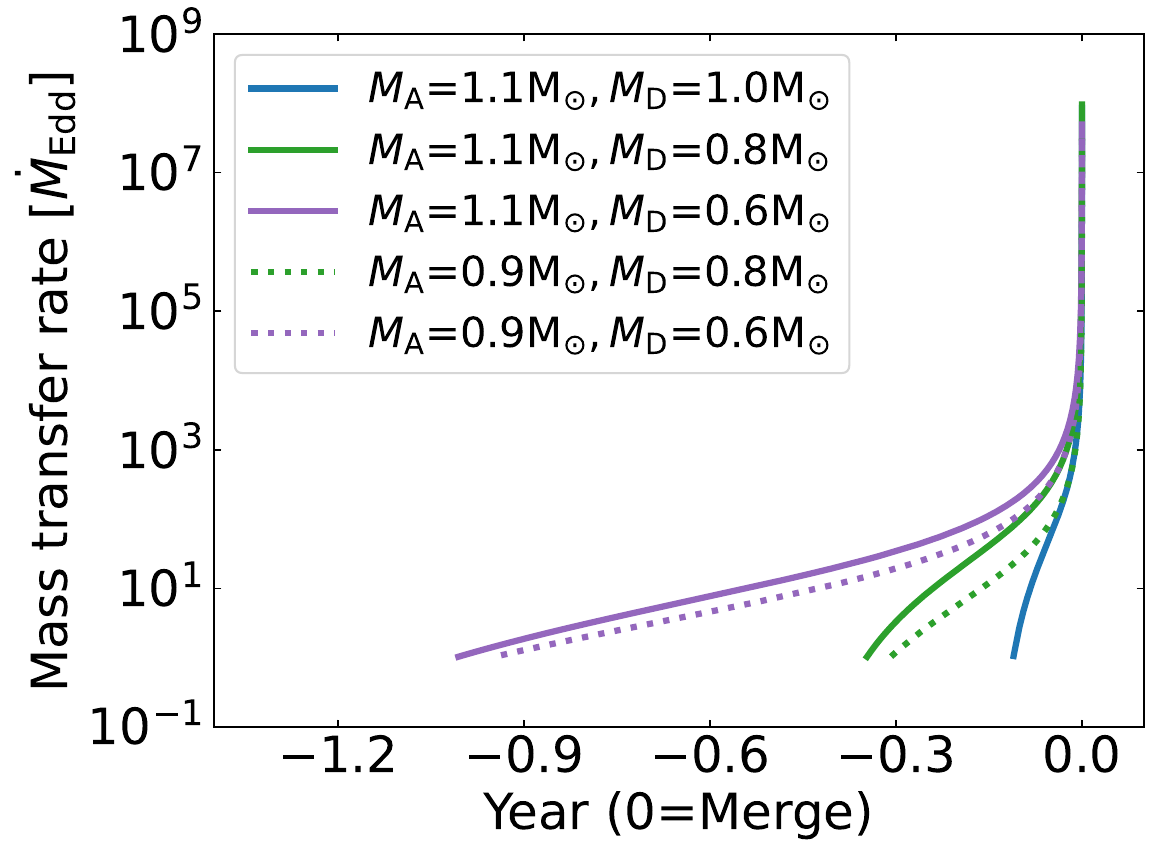}
        \end{minipage}
            \caption{The evolution of the orbital separations ($a$: left panels) and the mass-transfer rates ($\dot{M}_{\mathrm{D}}$ in unit of the Eddington's accretion limit: right panel). The solid and dotted lines are for $M_\mathrm{A}=1.1\ M_\mathrm{\odot}$ and $0.9\ M_\mathrm{\odot}$, respectively. Different line colors are used for different donor WD masses; $M_\mathrm{D}=1.0\ M_\mathrm{\odot}$ (blue), $0.8\ M_\mathrm{\odot}$ (green), and $0.6\ M_\mathrm{\odot}$ (purple). 
            The Eddington accretion rate is typically $\sim10^{-5}\ M_{\odot}\ \mathrm{yr^{-1}}$ in any models.
            \label{fig:orbitalevolution}}
        \end{figure*}

        The mass accretion rate to the accretor ($\dot{M}_{\mathrm{A}}$) takes into account the mass loss by the wind:
        \begin{equation}\label{eq:mdota}
            \begin{split}
                \dot{M}_{\mathrm{A}}
                =|\dot{M}_{\mathrm{D}}|-\dot{M}_{\mathrm{wind}},
            \end{split}
        \end{equation}
        where $\dot{M}_{\mathrm{wind}}$ is the mass loss rate.
        We assume that the wind is launched when the accretion rate exceeds the Eddington accretion rate, 
        \begin{equation} \dot{M}_{\mathrm{Edd}}=\frac{L_{\mathrm{Edd}}}{\phi_{\mathrm{L1}}- \phi_{\mathrm{A}}}\ ,
        \end{equation}
        where $L_{\rm Edd}$ is the Eddington luminosity, and $\phi_{\rm L1}$ and $\phi_{\rm A}$ are the gravitational potential energies at the inner Lagrangian point and the accretor's surface, respectively \citep{1999A&A...349L..17H}. 
        For the opacity, we simply adopt 0.2 cm${^2}$ g$^{-1}$ assuming Thomson scattering in hydrogen-poor matter.
        The Eddington accretion rate is $\sim10^{-5}\ M_{\odot}\ \mathrm{yr^{-1}}$ for our parameter sets (see Section \ref{sec:Results}), but it moderately changes as the potentials evolve during the mass transfer.\par

        We assume that the wind mass-loss rate is given by $\dot{M}_{\mathrm{wind}}=f_{\mathrm{loss}}|\dot{M}_{\mathrm{D}}|$, where the coefficient $f_{\rm loss}$ is obtained by the energy conservation \citep{1999A&A...349L..17H}, 
        \begin{equation}\label{eq:energyconswind}
            |\dot{M}_{\mathrm{D}}|\phi_{\mathrm{L1}}=(1-f_{\mathrm{loss}})|\dot{M}_{\mathrm{D}}|\phi_{\mathrm{A}}+f_{\mathrm{loss}}|\dot{M}_{\mathrm{D}}|\frac{v_{\mathrm{wind}}^{2}}{2}\ ,
        \end{equation}
        where $v_{\rm wind}$ is the terminal wind velocity, which is a free parameter in this study. 
        Then the coefficient is given by
        \begin{equation}\label{eq:floss}
            f_{\mathrm{loss}}=
            \frac{\phi_{\mathrm{L1}}-\phi_{\mathrm{A}}}{-\phi_{\mathrm{A}}+(v_{\mathrm{wind}}^{2}/2)}{\,\,\,\mathrm{for}\,\,\,}|\dot{M}_{\mathrm{D}}|>\dot{M}_{\mathrm{Edd}}\ ,
        \end{equation}
        and $f_{\rm loss}=0$ for $|\dot{M}_{\mathrm{D}}|\leq\dot{M}_{\mathrm{Edd}}$. 
        To estimate the potential energies, we assume that the accretor's radius is given by eq. \eqref{eq:massradiusrela} for simplicity. 
        Within a plausible range of the wind velocity, $v_{\rm wind}\sim10^8\text{--}10^9\,\rm cm\,s^{-1}$ (see Section \ref{subsec:csmmethod}), the wind coefficient is insensitive to $v_{\rm wind}$ and has a value of $f_{\rm loss}\sim0.1$ because of $|\phi_{\rm A}|\gtrsim v_{\rm wind}^2/2$.\par
        The evolution of the angular momentum change ($\dot{J}_{\mathrm{orb}}/J_{\mathrm{orb}}$) is mainly driven by gravitational wave \citep[$\dot{J}_{\mathrm{GW}}/J_{\mathrm{orb}}$:][]{Peters1964}, the wind \citep[$\dot{J}_{\mathrm{wind}}/J_{\mathrm{orb}}$: e.g.,][]{2008ApJ...679.1390H} from the binary system, and the spin up of the accretor by the mass transfer \citep[$\dot{J}_{\mathrm{MT}}/J_{\mathrm{orb}}$: e.g.,][]{2001A&A...368..939N,2004MNRAS.350..113M}.
        We assume that the WDs initially corotate with the binary orbit.
        Thus, $\dot{J}_{\mathrm{orb}}/J_{\mathrm{orb}}$ is expressed as follows:
        \begin{align}\label{eq:angularmomentumchange1}
            \frac{\dot{J}_{\mathrm{orb}}}{J_{\mathrm{orb}}}&=\frac{\dot{J}_{\mathrm{GW}}}{J_{\mathrm{orb}}}
            +\frac{\dot{J}_{\mathrm{wind}}}{J_{\mathrm{orb}}}
            +\frac{\dot{J}_{\mathrm{MT}}}{J_{\mathrm{orb}}},
        \end{align}
        where the three terms are given by
        \begin{align}
            &\frac{\dot{J}_{\mathrm{GW}}}{J_{\mathrm{orb}}}=-\frac{32}{5}\frac{G^{3}}{c^{5}}\frac{M_{\mathrm{A}}M_{\mathrm{D}}(M_{\mathrm{A}}+M_{\mathrm{D}})}{a^{4}}\label{eq:gw},\\
            &\frac{\dot{J}_{\mathrm{wind}}}{J_{\mathrm{orb}}}=-l_{\mathrm{wind}}\frac{M_{\mathrm{A}}+M_{\mathrm{D}}}{M_{\mathrm{A}}M_{\mathrm{D}}}\dot{M}_{\mathrm{wind}}\label{eq:wind},\\
            &\frac{\dot{J}_{\mathrm{MT}}}{J_{\mathrm{orb}}}=[(1+q)r_{\mathrm{h}}]^{0.5}\frac{\dot{M}_{\mathrm{D}}}{M_{\mathrm{D}}}\label{eq:mt}.
        \end{align}
        Here $l_{\mathrm{wind}}$ is the specific angular momentum of the wind in units of $a^{2}\Omega_{\mathrm{orb}}$, where $\Omega_{\mathrm{orb}}$ is the orbital angular velocity. 
        The value of $l_{\mathrm{wind}}$ is an order of unity \citep[e.g.,][]{1999ApJ...522..487H,2005A&A...441..589J,2020ApJ...895...29M}, but the exact value depends on the detailed situation, e.g., the acceleration process of the wind \citep[see][for the details]{2005A&A...441..589J}.
        We simply adopt $l_{\mathrm{wind}}=1$.\par
    
        The orbital radius of accretion material around the accreting WD, $r_{\mathrm{h}}$, is defined in unit of the orbital separation ($a$) in eq. \ref{eq:mt}. 
        For $r_{\mathrm{h}}$, we employ a fitting formula provided by \citet{2001A&A...368..939N} \citep[see also][]{1975ApJ...198..383L} expressed as $r_{\mathrm{h}}\simeq 0.04948-0.03815\mathrm{log}(q)+0.04752\mathrm{log}^{2}(q)-0.006973\mathrm{log}^{3}(q)$.
        In the present situations described in Sections \ref{sec:method} and \ref{sec:Results}, $r_{\mathrm{h}}a/R_{\mathrm{A}}\simeq0.3\le1$ (accretor radius $R_{\mathrm{A}}$ follows eq. \ref{eq:massradiusrela}), i.e., the direct impact accretion occurs instead of the disk accretion. 
        Thus, the accreting WD directly receives the angular momentum of the accreted matter. \par
        For the typical values of the present situation (e.g., $r_{\mathrm{h}}=0.05$, $q=1$, $1 M_{\odot}$ WDs and $a=10^{-2}\ R_{\odot}$), we see that the GW effect is $\sim10^{-9}\ \mathrm{s^{-1}}$ while the latter two effects are $\sim10^{-9}(\dot{M}_{\mathrm{D}}/0.1\ M_{\odot}\ \mathrm{yr^{-1}})\ \mathrm{s^{-1}}$. This exercise shows that 
        there is a critical value in $\dot{M}_{\mathrm{D}}$, $\sim 0.1\ M_{\odot}\ \mathrm{yr^{-1}}$, at which the main mechanism of the orbital evolution changes between the GW radiation and the mass transfer (see also Section \ref{sec:Results}).\par  
        For the numerical integration of eq. \ref{eq:orbitalevolution}, we conduct fifth-order Runge–Kutta integrations with an initial separation satisfying $|\dot{M}_{\mathrm{D}}|\leq \dot{M}_{\mathrm{Edd}}$ to capture the outer edge of the circumstellar wind. 
        We terminate the integration when the orbital separation shrinks below the tidal radius $a_{\mathrm{tidal}}=2(M_{\mathrm{A}}/M_{\mathrm{D}})^{1/3}R_{\mathrm{D}}$, at which the donor is tidally disrupted.

    \subsection{The CSM density distribution}\label{subsec:csmmethod}
        The wind driven by the super-Eddington mass transfer forms the CSM.
        Assuming spherical geometry, the radial distribution of the CSM density at the moment of SN explosion is expressed as follows:
        \begin{equation}\label{eq:rhocsm}
            \rho_{\mathrm{CSM}}(r)=\frac{\dot{M}_{\mathrm{wind}}}{4\pi r^{2} v_{\mathrm{wind}}}.
        \end{equation}
        Here, the mass-loss rate is evaluated as the one at $t$ before the tidal disruption for the wind material at $r=v_{\rm wind}(t+t_{\rm lag})$ under the assumption of a constant wind velocity. 
        \par
        The timescale, $t_{\mathrm{lag}}$, represents the time lag between the tidal disruption of the donor and the SN Ia explosion.
        In the present study, we simply fix $t_{\mathrm{lag}}$ as $100\ \mathrm{s}$; in the VM scenario, the primary WD explodes within the dynamical timescale after the onset of disruption of the donor \citep[e.g.,][]{2012MNRAS.424.2222P,2015ApJ...807..105S,2015ApJ...807...40T}, 
        \par
        For the terminal wind velocity ($v_{\mathrm{wind}}$), we explore three possibilities, $v_{\mathrm{wind}}=1000$, $4000$ and $8000\ \mathrm{km\ s^{-1}}$, because the wind acceleration mechanism is unclear.
        The first choice roughly corresponds to the asymptotic velocity of the matter ejected from the outer Lagrangian (L2) point, $\sim0.2(2G(M_{\mathrm{A}}+M_{\mathrm{D}})/a)^{0.5}$ \citep[][]{2016MNRAS.455.4351P}.
        The latter two roughly adopt the escape velocities of the accreting WD ($\sim(2GM_{\mathrm{A}}/R_{\mathrm{A}})^{0.5}$), and the DD system ($\sim(2G(M_{\mathrm{A}}+M_{\mathrm{D}})/a)^{0.5}$), respectively.\par

        We discuss the CSM density distribution in Section \ref{sec:Results}. Observational properties of the SN-CSM interaction are investigated in Sections \ref{sec:vmLC} and \ref{subsec:xray}.\par

    \begin{figure}[t]
    \centering
    \includegraphics[keepaspectratio, scale=0.6]{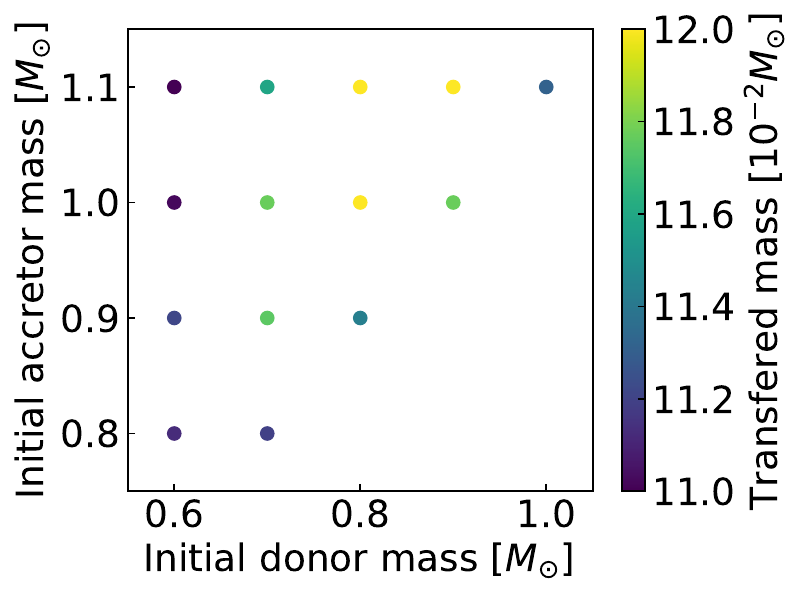}
        \caption{The transferred mass before the merger (i.e., when the separation reaches the tidal radius). 
        \label{fig:transmass}}
    \end{figure}
    
    \begin{figure}[t]
    \centering
    \includegraphics[keepaspectratio, scale=0.6]{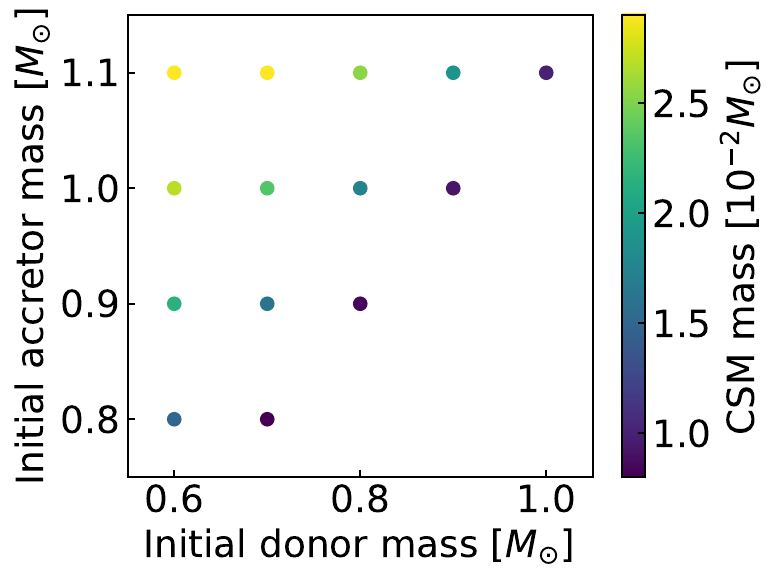}
        \caption{The CSM mass from by the evolution toward the merger. 
        \label{fig:csmmass}}
    \end{figure}
   
\section{Orbital evolution and CSM formation} \label{sec:Results}
    In this Section, we discuss the orbital evolution of double WD binaries and the CSM formation, assuming $v_{\mathrm{wind}}=4000\ \mathrm{km\ s^{-1}}$.
    We adopt different combinations of $M_\mathrm{A}$ ($0.9$ and $1.1\ M_\odot$) and $M_\mathrm{D}$ ($0.6$, 0.8 and $1.0\ M_\odot$). This allows us to explore a range of binary configurations, which encompasses the probable mass range for the VM scenario \citep[e.g.,][]{2011A&A...528A.117P,2015ApJ...807..105S,2016ApJ...821...67S} \footnote{According to \citet{2011A&A...528A.117P}, a violent merger (VM) may occur in double WD systems with the mass ratio of $q>0.8$. 
    However, simulations of VMs are highly sensitive to numerical resolution \citep[][]{2012MNRAS.422.2417D}, and thus suffer from large uncertainties.
    In the present work, we therefore adopt the conservative parameter space.}.
    Note that the He-ignited VM scenario, which is potentially realized in some WD-WD merger systems \citep[e.g.,][]{2010ApJ...709L..64G,2013ApJ...770L...8P}, is not considered here.
    It is discussed in Section \ref{subsec:he_vm}.\par
    The left panel of Figure. \ref{fig:orbitalevolution} shows the evolution of the orbital separation. 
    We find that the binary coalesces within $\sim0.1\text{--}1.0\ \mathrm{yr}$ after the onset of the super-Eddington mass transfer.
    The orbital separation remains almost unchanged until it starts shrinking rapidly in the last few weeks toward the merger.
    In the former long and slow evolution phase, the evolution is governed by gravitational wave radiation. In this phase, a small decrease in the orbital separation gradually increases the mass-transfer rate, as shown in the right panel of Figure. \ref{fig:orbitalevolution}.  Thus, gravitational wave radiation controls the duration of the super-Eddington mass transfer phase.
    As shown in Figure \ref{fig:orbitalevolution}, the lower mass-ratio binaries have a longer super-Eddington mass transfer phase.x
    \par

    Once the mass transfer rate reaches $\sim10^4\,\dot{M}_{\rm Edd}(\sim0.1~ M_{\odot}\rm ~ yr^{-1})$, the mass transfer and wind begin to dominate the angular momentum evolution, causing the system to evolve much more rapidly. Given that the shrinking of the orbital separation is mostly driven in this phase by the mass transfer and wind, the total transferred and ejected masses are mainly provided in this phase. 
    Figure \ref{fig:transmass} shows that the transferred mass given by the integration of $\dot{M}_{D}$ during the numerical integration is almost identical, $\sim0.1\ M_{\odot}$, in any models. 
    Based on eq. \ref{eq:orbitalevolution}, the transferred mass is roughly determined by the difference between initial separation and tidal radius; while models with smaller mass ratios have larger initial separations ($a_{\rm init}\propto M_{\rm D}^{-2/3}$), they have larger tidal radii ($a_{\rm tidal}\propto M_{\rm D}^{-1/3}$). Consequently, the transferred mass does not monotonically increase with smaller $M_{D}$ for a given $M_{A}$.
    Figure \ref{fig:csmmass} shows that the CSM mass is $\sim0.01\text{--}0.03\ M_{\odot}$, and thus we find that $\sim10\%$ of the transferred mass is ejected during the super-Eddington mass transfer phase.
    In addition, we find that the smaller mass-ratio binary models have larger CSM mass. 
    This is because $f_{\mathrm{loss}}$ is larger for larger $\phi_{\mathrm{L1}}- \phi_{\mathrm{A}}$, i.e., in the smaller mass-ratio binary models. 
    While tidal disruption also contributes to the total CSM mass, its effect on our main conclusions is negligible, as discussed in Section \ref{subsec:lines}.
    \par

\begin{figure}[t]
    \centering
    \includegraphics[keepaspectratio, scale=0.3]{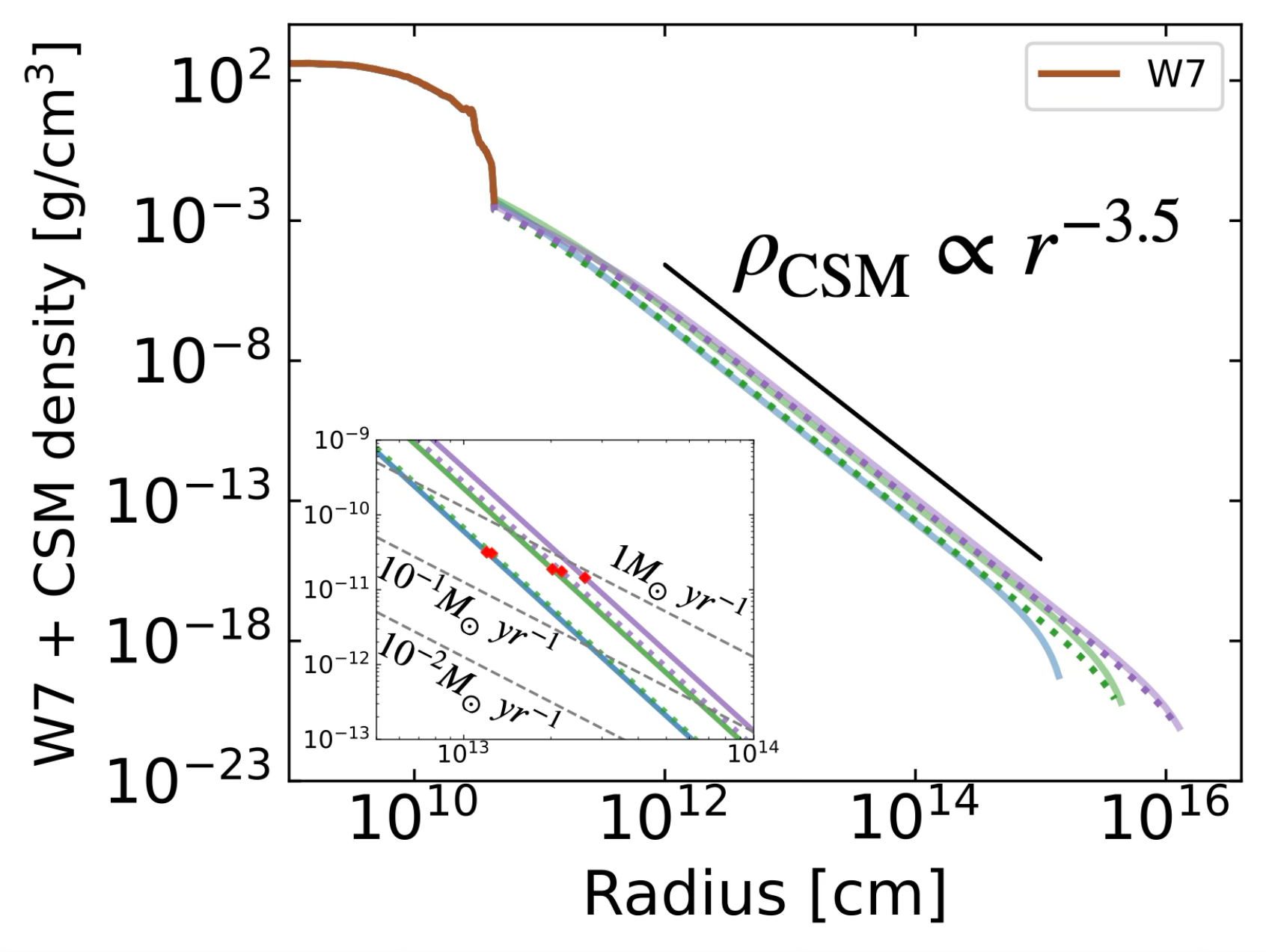}
        \caption{
        The CSM density distribution for the same set of the models from Figure \ref{fig:orbitalevolution}. Also shown is the W7 ejecta model (brawn).
        The black solid line shows the slope of $r^{-3.5}$. 
        The inset is an expanded view at $5\times10^{12}\text{--}10^{14}$ cm.
        The red points represent the shock-breakout radii, corresponding to the optical depth of 30 measured from the outside (i.e., the shock velocity is assumed to be $10000\ \mathrm{km\ s^{-1}}$). Thomson scattering in hydrogen-poor matter, $0.2\ \mathrm{cm^{2}\ g^{-1}}$, is used for the opacity. 
        The CSM density structure for the steady-state mass loss is given by the grey lines (assuming a constant velocity of $v_{\mathrm{wind}}=4000\ \mathrm{km\ s^{-1}}$).
        \label{fig:csmproperty}}
    \end{figure}

    Figure \ref{fig:csmproperty} shows that the CSM density distribution roughly follows $r^{-3.5}$, irrespective of the model parameters. 
    This steep (and universal) density distribution reflects the rapid increase in the mass-loss rate  during the final phase when gravitational wave radiation can be ignored (see Appendix \ref{app:csmslope}). The inset of Figure \ref{fig:csmproperty} shows an expanded view around the shock breakout radius following the SN-CSM interaction (Section \ref{sec:vmLC}). 
    We find that the differences in the CSM density scale for different binary parameters are within an order of magnitude; $\rho_{\mathrm{CSM}}\simeq D(r/10^{14}\ \mathrm{cm})^{-3.5}\ (D\simeq 10^{-14}\text{--}10^{-13}\ \mathrm{g\ cm^{-3}})$, where 
    a smaller mass-ratio model tends to have a denser CSM because of the dependence of $f_{\mathrm{loss}}$ on the mass ratio (see above). 
    The locations of the CSM outer edges are also within an order of magnitude, because it is controlled by the duration of the super-Eddington mass transfer phase which spans $\sim 0.1\text{--}1$ year for different models; for a lower mass-ratio model, a more extended CSM distribution is expected because of the longer super-Eddington mass transfer phase. 

    \begin{figure}[t]
    \centering
    \includegraphics[keepaspectratio, scale=0.55]{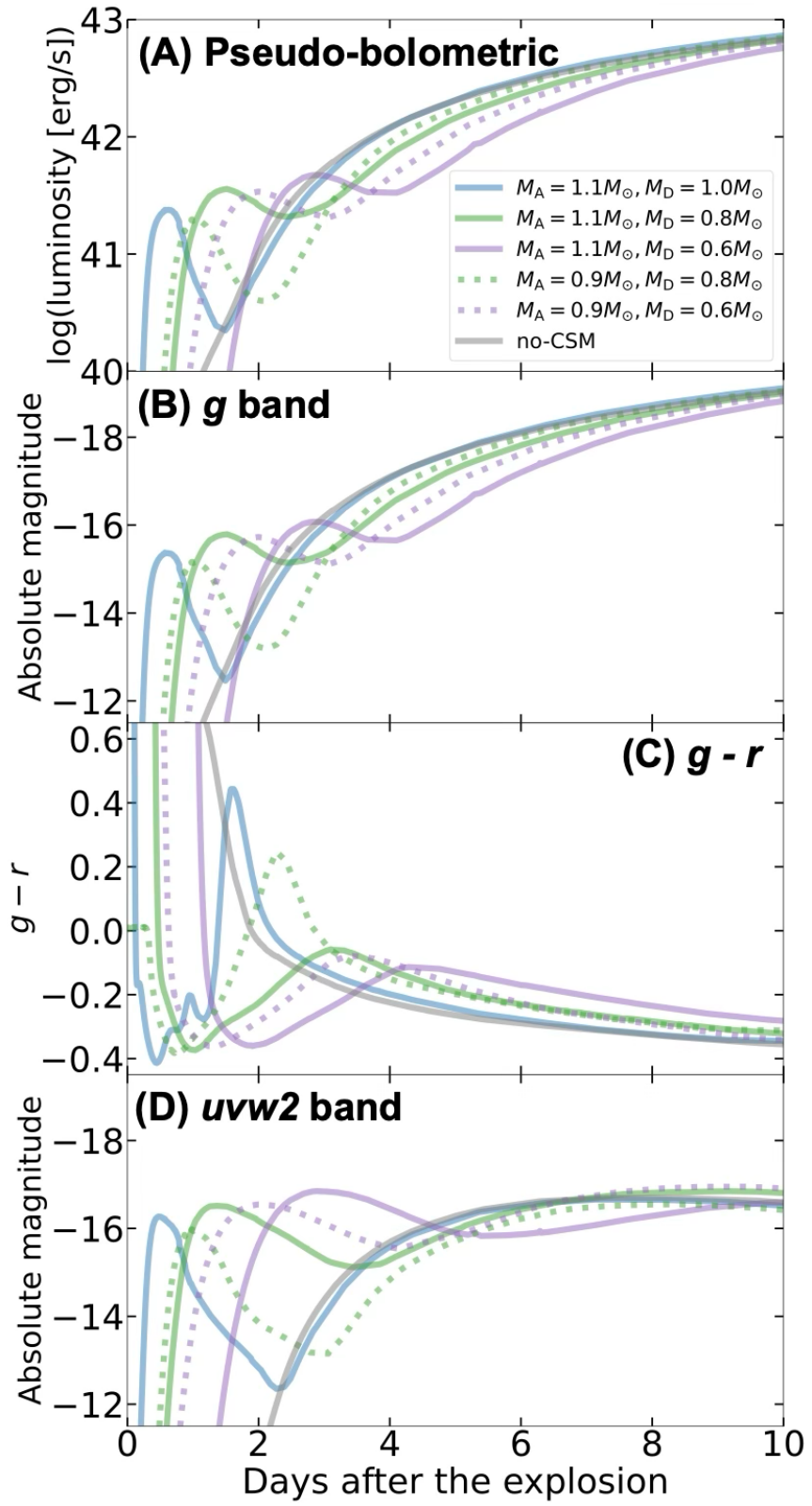}
        \caption{The synthesized photometric evolutions for the same set of models from Figure \ref{fig:orbitalevolution}. Also shown is the no-CSM model (grey). Shown here are (A) the pseudo-bolometric LCs (3250-8900 {\AA}), (B) the $g$-band LCs, (C) the $g-r$ color evolution, and (D) the $uvw2$-band LCs. 
        }
        \label{fig:synthesizedLCs}
    \end{figure}

\section{Photometric signatures of the CSM interaction}\label{sec:vmLC}
    To test our CSM formation scenario, we perform LC calculations for the interaction between the SN Ia ejecta and the CSM.
    We use the AB magnitude system in the present study.
    \subsection{Setup for the hydrodynamic and light curve simulations}\label{subsec:LCmethod}
    We calculate the LCs using 1-dimentional (1D) radiation-hydrodynamic code, STELLA \citep{1998ApJ...496..454B,2000ApJ...532.1132B,2006A&A...453..229B}, which is available as a part of MESA version 24.08.1 \citep{Paxton2011,Paxton2013,Paxton2015,Paxton2018,Paxton2019,Jermyn2023}.
    For the SN ejecta, we adopt the W7 model \citep[see Figure \ref{fig:csmproperty};][]{1984ApJ...286..644N,1986A&A...158...17T} as a proxy, noting that the long-term LC evolution , which can be sensitive to details of the explosion mechanism, is not a focus of the present work. 
    The CSM density distribution is taken from the result of our binary evolution model (see Figure \ref{fig:csmproperty}). For comparison, we also perform the LC calculation based on the W7 model without attaching the CSM (`no-CSM model'). 
    We assume that the composition of the CSM consists of equal mass fractions of carbon (C) and oxygen (O).

    \subsection{LC and color evolution in the optical and UV}\label{subsec:LCresult}
    We discuss the LC properties across the optical to UV bands using the models with $v_{\mathrm{wind}}=4000\ \mathrm{km\ s^{-1}}$ as our reference models.
    The synthesized pseudo-bolometric LCs, which are obtained by integrating all the fluxes in the optical wavelengths between 3250 and 8900 {\AA}, are shown in Figure \ref{fig:synthesizedLCs}(A). 
    The cases with the CSM show early flux excesses, compared to the no-CSM model.
    We find that the flux excesses emerge within a few days of the explosion.
    Since the shock breakout radii are located at $\sim1\text{--}3\times10^{13}\ \mathrm{cm}$ (see Figure \ref{fig:csmproperty}) and the velocity of the SN ejecta is $\simeq10000\ \mathrm{km\ s^{-1}}$, the shock breakout occurs within $\sim0.1$ days after the explosion.
    On the other hand, the peaks of the flux excesses occur within a few days of the explosion.
    As such, the early flux excess here is analogous to the `cooling-envelope' emission discussed mainly for core-collapse SNe with an extended envelope \citep[e.g.,][]{2016ApJ...826...96P,2018ApJ...861...78M}. 
    After the early flux excess, the luminosity asymptotically approaches that of the no-CSM model.\par
    The synthesized LCs in the $g$-band of the Zwicky Transient Facility \citep[ZTF;][]{2019PASP..131a8002B} are shown in Figure \ref{fig:synthesizedLCs}(B). 
    We find that the absolute $g$-band magnitudes of the peaks of the excesses are $\sim - 15\text{--}-16$ mag.
    In addition, the synthesized color ($g-r$) evolution is shown in Figure \ref{fig:synthesizedLCs}(C).
    All the models start with an extremely red color, and evolve toward the blue as photons diffuse out of the ejecta (and the CSM). This transition from the red to blue takes place much faster in the models with the CSM following the wind shock breakout; the photons created by the SN-CSM interaction have much shorter diffusion time scale than those created by the $^{56}$Ni decay inside the ejecta. These models rapidly reach to $g-r\simeq-0.4$, and then move back to the red, finally (roughly) following the no-CSM model as dominated by the $^{56}$Ni heating. The timescale for this behavior is faster for the models with lower CSM density, i.e., the higher binary mass-ratio models (Figure \ref{fig:csmproperty}).
    \par
    The STELLA simulations show a strong UV early-excess emission in the $uvw2$-band of the Neil Gehrels Swift Observatory \citep[][]{2004ApJ...611.1005G,2005SSRv..120...95R}.
    The peaks ($\sim-16\text{--}-17$ mag) are brighter than in the $g$-band.
    The durations of the UV excesses are $\sim0.5\text{--}4$ days, which are slightly longer than those of the optical excesses.
    This strong emission results from a high blackbody temperature of $\sim50000\ \mathrm{K}$.
    \citet{2018ApJ...861...78M} estimated a typical blackbody temperature of the early excess as $\sim7000\text{--}15000$ K in the He-detonation scenario and $\sim20000\text{--}25000$ K in the companion interaction scenario, which are much lower than that of our CSM interaction model.
    Thus, the high-cadence UV observation during the early-excess phase offers a great probe for the CSM interaction scenario.
    \par
\begin{figure*}[t]
        \begin{minipage}[b]{0.1\linewidth}
        \centering
        \includegraphics[keepaspectratio, scale=0.37]{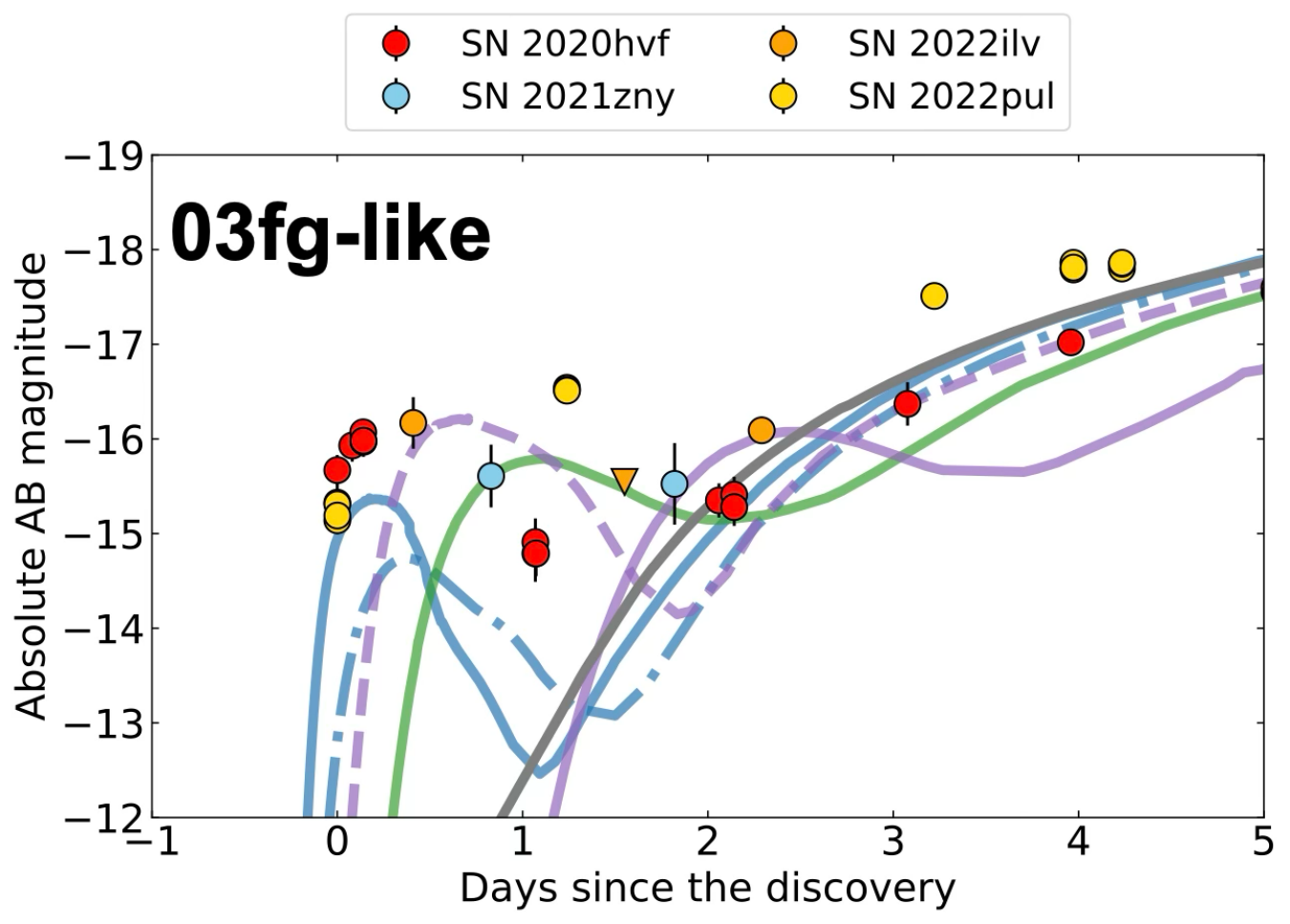}
        \end{minipage}
        \hfill
        \begin{minipage}[b]{0.5\linewidth}
        \centering
        \includegraphics[keepaspectratio, scale=0.37]{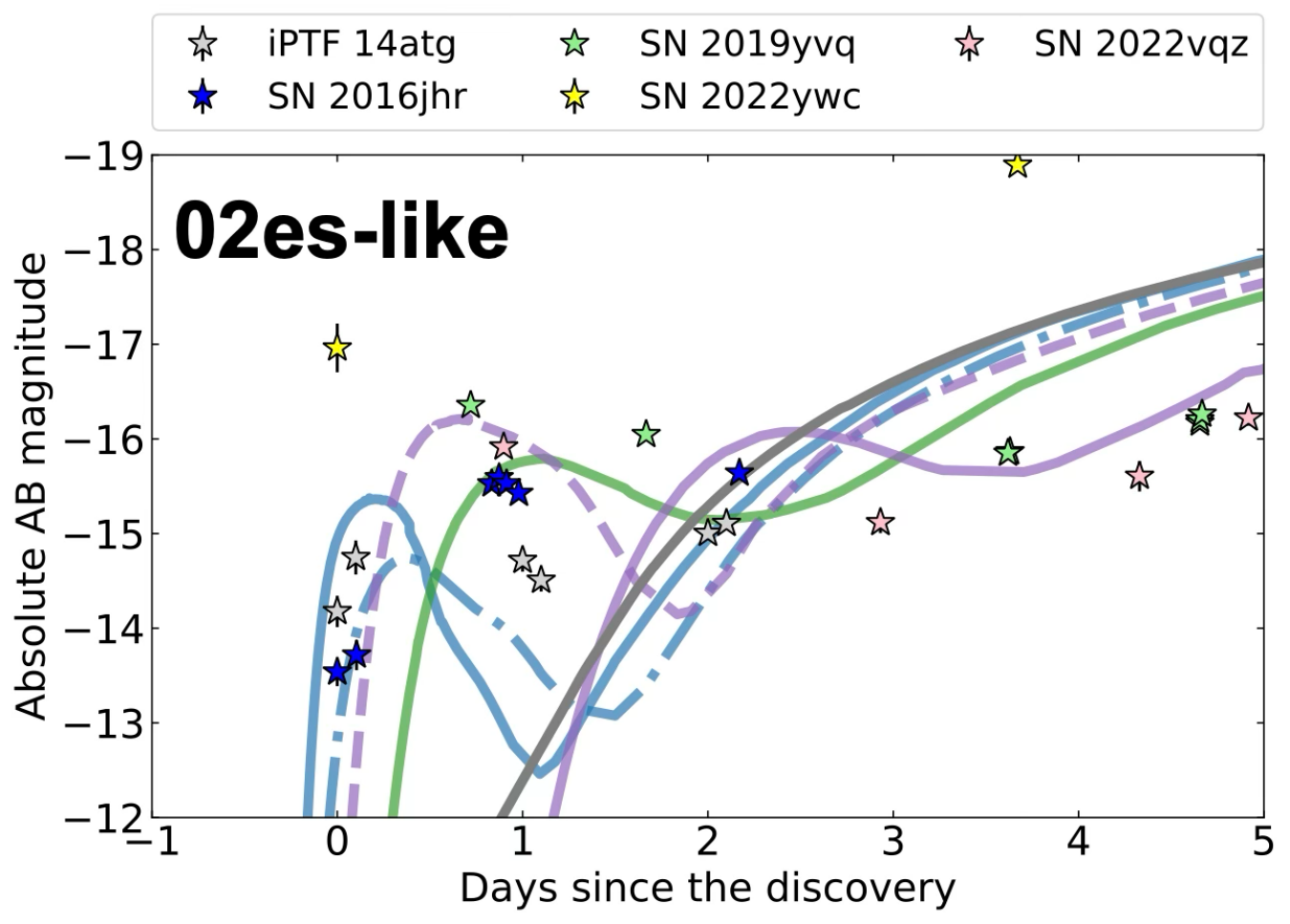}
        \end{minipage}
    \caption{
    The synthesized $g$-band LCs, as compared with the 03fg-like (left) and the 02es-like (right) objects (see Table \ref{table:obsdata}). 
    The models shown here are for $M_\mathrm{D}=1.0\ M_\mathrm{\odot}$ (blue), $0.8\ M_\mathrm{\odot}$ (green), and $0.6\ M_\mathrm{\odot}$ (purple) (see Figure \ref{fig:orbitalevolution}), while 
    the accretor mass is fixed to be $1.1\ M_{\odot}$.
    Different line styles are for different wind velocities; $v_{\mathrm{wind}}=1000\ \mathrm{km\ s^{-1}}$ (dashed), $4000\ \mathrm{km\ s^{-1}}$ (solid), and $8000\ \mathrm{km\ s^{-1}}$ (dashdot).
    For comaprison, the no-CSM model is also shown (grey). 
    In this figure, the explosion dates of the synthesized LCs are set to be $0.4$ days before the discovery.
    \label{fig:gband_obs}}
    \end{figure*}

    The synthesized $g$-band LCs of the models with $v_{\mathrm{wind}}=1000$, $4000$ and $8000\ \mathrm{km\ s^{-1}}$ (see Section \ref{subsec:csmmethod}) are shown in Figure \ref{fig:gband_obs}.
    For higher $v_{\mathrm{wind}}$, the peak in the early flux excess is reached at a later epoch. This stems from the higher CSM density for higher $v_{\mathrm{wind}}$ at a given radius (i.e., at a given observational epoch) for the following reason. 
    In general, CSM density profile follows $\rho_{\rm CSM}\propto \dot{M}_{\rm wind}/v_{\rm wind}$ at a given radius.
    In our model, the rapid increasing mass-loss rate shows the relation of $\dot{M}_{\rm wind} \propto \widetilde{t}^{-1.5} \propto v_{\rm wind}^{1.5}$, where $r\simeq v_{\rm wind} \widetilde{t}$ (see Appendix \ref{app:csmslope} for details).
    This leads to $\rho_{\rm CSM}\propto v_{\rm wind}^{0.5}$, i.e., the CSM density becomes higher for higher $v_{\mathrm{wind}}$ models.\par

    \subsection{Comparison to observational data} \label{subsec:obsdata}
\begin{figure}[t]
    \centering
    \includegraphics[keepaspectratio, scale=0.37]{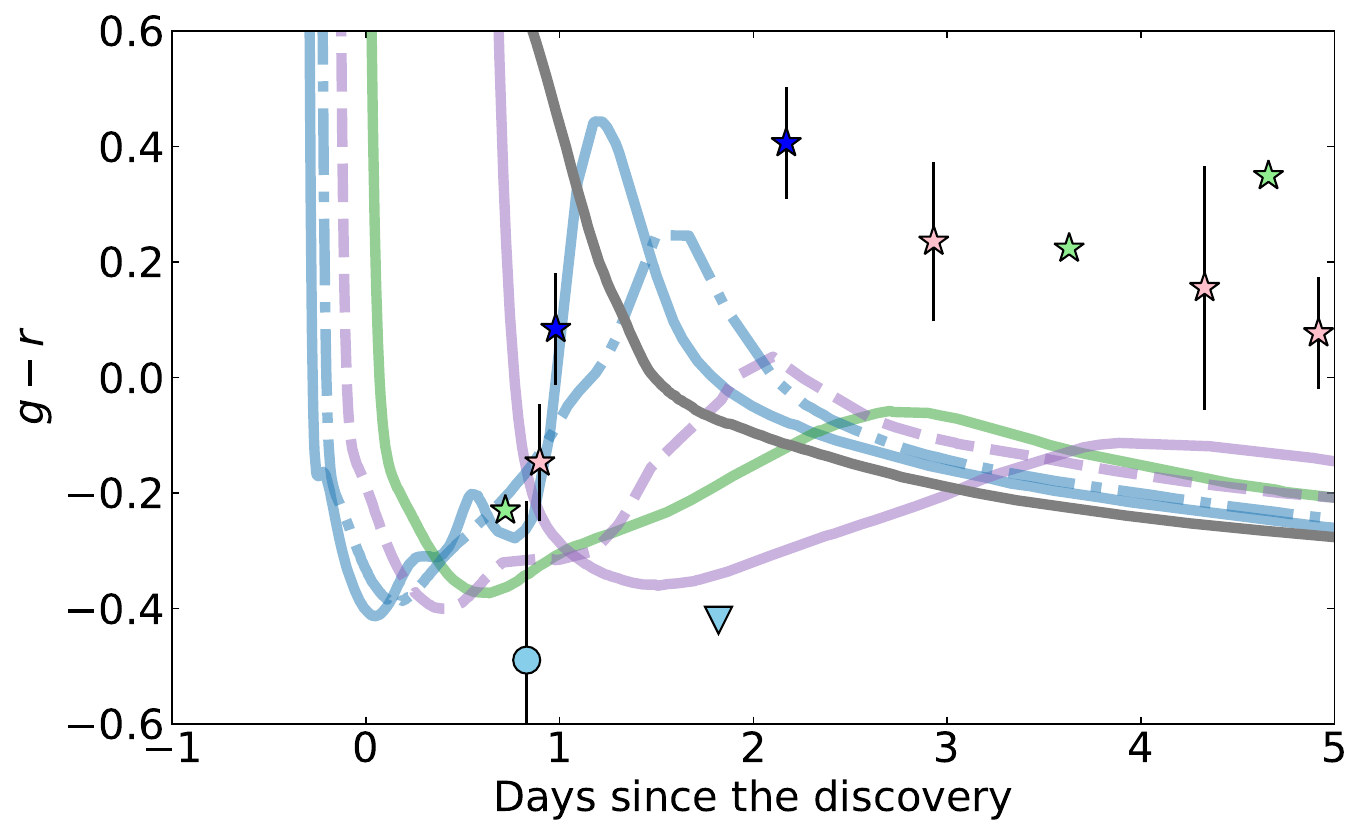}
        \caption{The synthesized and observed $g-r$ evolution. See the caption of Figure \ref{fig:gband_obs} for details of the models and data.
        \label{fig:g-r_obs}}
    \end{figure}

    Figure \ref{fig:gband_obs} compares the synthesized $g$-band LCs to the $g$-band LCs of 03fg- and 02es-like objects (see Table \ref{table:obsdata}).
    The observed LCs show that the peak magnitudes of the early excesses are $-15\text{--}-17\ \rm mag$, and the durations are $\sim1\text{--}2\ \rm days$.
    The synthesized LCs qualitatively match the observed early flux excesses in their brightness and durations, and also explain the variety of the early flux excesses 
    Furthermore, as shown in Figure \ref{fig:g-r_obs}, the synthesized color evolution qualitatively explains the observed blue color ($g-r\simeq-0.5\text{--}-0.2$) within $\sim1$ day after the discovery\footnote{While the observed color evolution after the early excess is different from the synthesized one, 
    this discrepancy likely stems from a systematic effect related to the W7 model \citep{2023MNRAS.522.6035M}.}.\par
    The exceptions are the very bright and/or long-duration early excesses seen in SNe 2022pul and 2022ywc. They require a more massive and extended CSM in the context of the CSM interaction scenario. Updating our model, e.g., including the CSM provided by the tidally disrupted debris, is needed to further test the VM scenario for these objects.\par
    UV observations within a few days after the SN explosion are still rare, but have been conducted for iPTF14atg \citep[at $\gsim 3$ days:][]{2015Natur.521..328C} and SN 2019yvq \citep[at $\gsim 1$ day:][]{2020ApJ...898...56M}.
    Their peak absolute magnitudes of the early flux excesses are $\sim-14.5$ mag and $\sim-16$ mag in the UV bands, being consistent with our CSM model (Figure \ref{fig:synthesizedLCs}(D)).\par
    In the context of the VM scenario, the lower peak luminosity of 02es-like objects than 03fg-like objects suggests that the accreting (and thus exploding) WD is less massive in the former \citep[e.g.,][]{2013ApJ...770L...8P}.
    On the other hand, the CSM density is largely independent of the mass of the accretor, but is primarily determined by the mass ratio (Figures \ref{fig:csmproperty} and \ref{fig:synthesizedLCs}).
    Therefore, no significant differences in the CSM density are expected between the 03fg- and 02es-like objects in this scenario.
    This prediction can be tested by increasing the sample of such objects observed in the earliest phase.
    \par
    We note that several mechanisms have been proposed for the early flux excess. For example, \citet{2017Natur.550...80J} argued that there are signs of He detonation in 02es-like objects, SN 2016jhr, e.g. deep $\rm Ti_{II}$ absorption lines. It indicates that some 02es-like objects may show the early excesses powered not by the CSM interaction, but by the radioactive decay of the He-detonation ash. We further note that the He-detonation-driven SNe Ia are also expected to form the CSM and the resulting SN-CSM interaction (Section \ref{subsec:he_vm}), and thus the early flux excess might be powered by multiple mechanisms in the He-detonation-driven SN Ia scenario. 
    To distinguish between the CSM-interaction and the He-detonation ash as a power source of the early excess, UV \citep[e.g.,][]{2024ApJ...966..139H,2025MNRAS.542.2752B} and X-ray observations are crucial (see Section \ref{subsec:xray}).
    For example, \citet{2020ApJ...898...56M} have shown that the peak UV brightness predicted by their He-detonation model is fainter than that of SN 2019yvq, while the model remains consistent with the optical brightness.

\section{Discussion}\label{sec:dis}
\subsection{Emission and absorption lines from CSM}\label{subsec:lines}
We suggest that searching for carbon and oxygen emission lines during the early excess would provide powerful diagnostics on the SN-CSM interaction scenario, once such an observation would be performed for 03fg/02es-like objects. This claim is constructed upon the following phenomenological analyses based on comparison to the situation for `SNe Icn' for which numerous carbon lines have been detected within several days after the explosion. We postpone detailed simulations of the spectral formation to the future. \par
SNe Icn are believed to be powered by the interaction between the SN ejecta and carbon-rich CSM. As a prototypical example, SN Ic 2019hgp had the peak bolometric luminosity of $\simeq \text{a few}\times 10^{43}\ \rm erg\ s^{-1}$ \citep[e.g.,][]{2022Natur.601..201G}. Its LC has been reproduced by the CSM density profiles following $\simeq 10^{-14}(r/5\times 10^{14}\ \rm cm)^{-3}\ g\ cm^{-3}$ \citep[][]{maeda2022ApJ...927...25M,2023A&A...673A..27N}, which translates into the CSM density at the shock front being higher than $10^{-14}\ \rm g\ cm^{-3}$ within several days since the explosion. These could be taken as conditions for strong carbon lines to emerge. \par
In our scenario for 03fg/02es-like objects, the CSM density at the shock front is as high as the above-mentioned value, at the peak of the early excess, i.e., $\sim 1$ days since the explosion (Figure \ref{fig:gband_obs}). The bolometric luminosity there is $\simeq 10^{43-44}\ \rm erg\ s^{-1}$, comparable to the case for SNe Icn. Indeed, the present model predicts a very high temperature ($T\sim 50000\ \rm K$) and the bolometric luminosity is contributed substantially by UV photons, providing conditions for photoionizations. These suggest similar ionization/thermal conditions in the pre-shocked CSM might be shared between the peak phase of SNe Icn and the early-excess phase of 03fg/02es-like objects. Note that the appearance of the very-early spectra migh well be different; the possible emission lines in the early-excess phase of 03fg/02es-like objects will have the width corresponding to $\sim 1000\text{--}10000$ km s$^{-1}$, which is much broader than seen in SNe Icn \citep[up to a few $1000$ km s$^{-1}$;][]{2022Natur.601..201G,2022ApJ...927..180P}.\par

Carbon absorption lines also provide different diagnostics. 
One of the distinguishing features of 03fg-like objects is strong carbon absorption lines in their spectra. It has been suggested that these lines might originate in the carbon-rich CSM (or envelope) swept up by the SN ejecta through the SN-CSM interaction; \citet{2023MNRAS.521.1897M} have shown that the  $\sim 0.01\text{--}0.1\ M_{\odot}$ of the carbon-rich CSM explains the properties of carbon absorption lines seen in 03fg-like objects. Our CSM formation model predicts the CSM mass of $\sim 0.01\text{--}0.03\ M_{\odot}$ that overlaps with this requirement, and thus the formation of carbon absorption lines in the post-excess phases is expected. 

We note that the CSM mass resulting from our simulations is not sufficient to explain extremely strong and persistent carbon lines observed in a fraction of 03fg-like objects (which require $\sim 0.1\ M_\odot$). This discrepancy could be resolved if an additional CSM (or envelope) mass would be provided by the tidally disrupted donor WD, i.e., the phase not considered in the present work (as we stopped our simulations at the onset of the tidal disruption). Note that this additional material that might exist in the innermost region of the CSM (or as an envelope surrounding the exploding WD) would not sensitively affect the nature of the early excesses; the additional interaction power provided there is likely lost by the adiabatic expansion cooling rather than powering the LCs. 

\subsection{X-ray as a smoking gun of our CSM model?}\label{subsec:xray}
In the context of the C/O-rich CSM interaction, the time evolution of emission at $\sim10$ keV could be a good tracer of the density distribution \citep[e.g.,][]{2025ApJ...980...86I}.

The peak time can be estimated by the date at which the optical depth ($\tau_{\mathrm{X}}$) of the unshocked CSM for X-rays becomes unity ($t_{\tau_{\mathrm{X}}=1}$).
By using $\rho_{\mathrm{CSM}}(r)\simeq D(r/10^{14}\ \mathrm{cm})^{-3.5}\ (D\simeq 10^{-14}\text{--}10^{-13} \mathrm{\ g\ cm^{-3}}$; Section \ref{sec:Results}) and the opacity for X-rays ($\kappa_{\mathrm{X}} \sim 3$ cm$^2$ g$^{-1}$ for the partially-ionized C/O-rich CSM), we estimate the peak time of the 10 keV emission as $\sim 2.5 \times (D/10^{-13})^{1/2.5}\ \mathrm{day}$. Here, we assumed the shock velocity of $10000~ \rm km~ s^{-1}$. 

\par
Free-free (FF) emission is expected to dominate the X-ray emission at 10 keV, provided that thermal electrons reach equilibrium with thermal ions at the shock front \citep{2006A&A...449..171N,2006ApJ...641.1029C,2006ApJ...651..381C}. 
In our case, the ion-electron equilibration timescale is shorter than the dynamical timescale in the initial phase just after the peak date \citep[e.g.,][]{2006ApJ...651..381C,maeda2022ApJ...927...25M}. 
In this situation, the cooling is dominated by the adiabatic cooling, since the shock accelerates as it propagates through a steep CSM density profile (steeper than $\rho_{\rm CSM}\propto r^{-3}$); 
this was confirmed by checking the hydrodynamical evolution following the method described by \citet{2025ApJ...980...86I}. We can thus estimate the luminosity by only considering the FF emission created immediately at the shock front. 
The peak luminosity is roughly proportional to $\rho_{\rm CSM}^{2} r_{\tau_{\mathrm{X}}=1}^3\propto D^{2} t_{\tau_{\mathrm{X}}}^{-4}\propto D^{0.4}$, and thus it is relatively insensitive to the exact value of $D$. We thus fix the value of $D$ as $10^{-13}$ g cm$^{-3}$ in the following discussion.
\par

The X-ray LC due to the FF emission at $10\ \mathrm{keV}$ is given as follows \citep{2006ApJ...641.1029C}: 
\begin{align}\label{eq:ffxrayflux}
        \frac{dL_{\mathrm{10\ \mathrm{keV}}}}{dE}
        \simeq2\times10^{42}\left(\frac{t}{3\ \mathrm{day}}\right)^{-4}\ \mathrm{erg\ s^{-1}\ keV^{-1}},
\end{align}
where we assume that the electron temperature is $10^{9}\ \mathrm{K}$ and the relative thickness of the forward shock region is 0.3 \citep[][]{1994ApJ...420..268C}. 
This X-ray is a clear signature that can be used to discriminate the CSM interaction scenario from other mechanisms such as the one involving the radioactive power from the He-detonation ash. 
This behavior of $t^{-4}$ could further provide evidence of the steep CSM density slope such as $\rho_{\rm CSM} \propto r^{-3.5}$ (see Figure \ref{fig:csmproperty} and Appendix \ref{app:csmslope}).

\par
At a distance of $100\ \mathrm{Mpc}$, the 10 keV X-ray flux is predicted to decrease from $\sim 10^{-12}$ to $10^{-13}\ \mathrm{erg\ s^{-1}\ keV^{-1}\ cm^{-2}}$ over $\simeq3\text{--}6$ days.
Such an event is a good target of XMM-Newton EPIC \citep{2001A&A...365L..51W} with an exposure time of $\sim10^3\ \rm s$.\par

\subsection{Geometry of the CSM}\label{subsec:csmgeo}
In this study, the CSM is treated under the spherical-symmetry assumption. In reality, the outflow from a binary system might create asymmetric CSM, e.g., disk-like CSM concentrated on the binary's orbital plane as suggested for 03fg-like objects \citep{2023MNRAS.521.1897M}.
If we observe the SN-disk CSM interaction \citep[e.g.,][]{2019ApJ...887..249S} from the face-on angle, the region of the CSM interaction might be hidden by the SN ejecta, resulting in substantial suppression in the early excess \citep{nagao2020MNRAS.497.5395N}. 

\par
Such a viewing-angle effect would lead to diversity in the properties of the early excess, which could be used as a test for the present scenario with an increasing number of 03fg- and 02es-like objects observed from the first few days of the explosion. For example, the double-detonation has been sometimes suggested for 02es-like objects, in which the early excess could be created by the radioactive decay of the He-detonation ashes \citep{2017Natur.550...80J,2018ApJ...861...78M}. It is expected that the viewing-angle effects are different; disk-like CSM interaction likely creates a stronger dependence on the viewing angle \citep[e.g.,][]{nagao2020MNRAS.497.5395N}. Details will depend on the exact configuration, and thus further theoretical study will help constrain the mechanism.

\subsection{A case of the He-ignited VM}\label{subsec:he_vm}
Prompt SN Ia explosions in WD-WD merger systems can also be triggered by the He-detonation, in the so-called He-ignited VM or D6 (dynamically-driven double-degenerate double-detonation) model \citep[e.g.,][]{2013ApJ...770L...8P,2018ApJ...868...90T,2019ApJ...885..103T,2022MNRAS.517.5260P}. The He detonation may be triggered more easily than the carbon detonation -- a question then is, given that there is a sufficiently large amount of He on the surfaces of the two WDs, whether the conditions for double detonation are satisfied before the onset of tidal disruption of the donor. 
\par
\begin{figure}[t]
\centering
\includegraphics[keepaspectratio, scale=0.37]{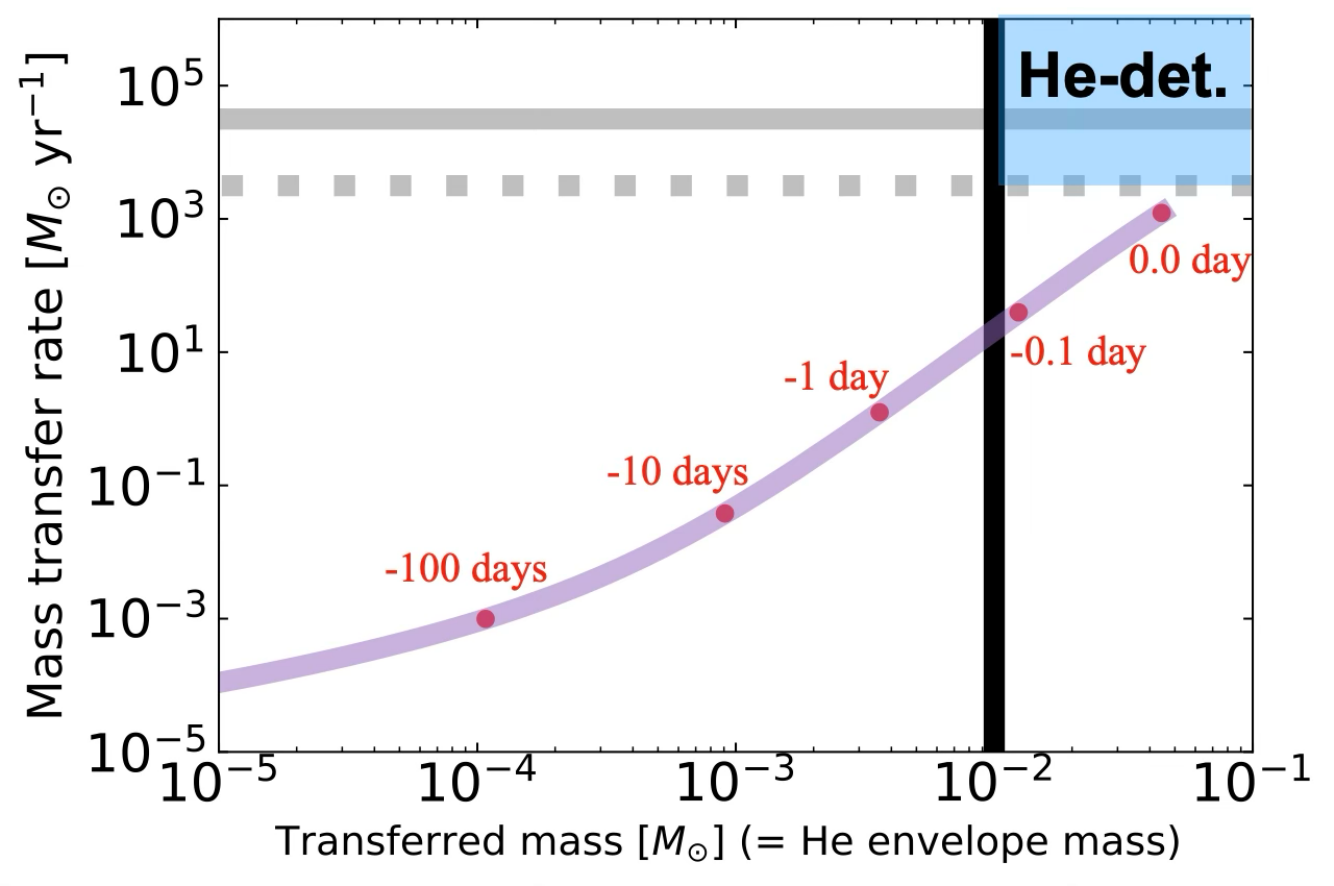}
    \caption{The temporal evolution of the mass transfer rate and the transferred mass.
    The shown here (purple solid line) is a binary model of $1.1\ M_{\odot}+0.6\ M_{\odot}$.
    The transferred mass represents the lower limit for the He-envelope mass of the primary star, given that the primary WD might already have the He envelope before the mass transfer \citep[$\sim10^{-3}\ M_{\odot}$ for a $1.1\ M_{\odot}$ WD, and $\sim10^{-2}\ M_{\odot}$ for a $0.6\ M_{\odot}$ WD;][]{2024ApJ...975..127S}.
    The labels shown in red denote the time before the onset of the tidal disruption of the donor.
    The He-ignition region is shaded by a blue area.
    \label{fig:he_vm}}
\end{figure}

We adopt the ignition conditions as follows \citep{2022ApJ...941...87I,2025ApJ...979...54R}: the mass transfer rate of $\gtrsim 10^{-4}\text{--}10^{-3}\ \rm M_{\odot}\ s^{-1}$ and the He-envelope mass of $\gtrsim 0.01\ M_{\odot}$.
Figure \ref{fig:he_vm} shows that our WD-WD binary model does not reach the ignition criteria before the onset of the tidal disruption of the donor. 

Therefore, the He-ignited VM will share the properties of the CSM with the C-ignited VM, i.e., the CSM formed by the pre-tidal disruption mass transfer phase. 
We thus suggest that the He-ignited VM should also be accompanied by the 03fg/02es-like early excesses through the CSM interaction (which might be further contaminated by the radioactive decay power). As possible diagnostics between the He- and C-ignited VMs, investigating spectral features in the first 1 day might be an interesting possibility (Section \ref{subsec:lines}) -- here, we may see He lines instead of carbon lines.  
\par

\subsection{Hyper velocity stars}\label{subsec:hvs}
In C- and He-ignited VMs, a WD-WD binary can eject the donor star as a hyper velocity stars (HVS) with a velocity of $\sim1000\ \rm km\ s^{-1}$ if the donor WD survives \citep[e.g.,][]{2018ApJ...865...15S,2025arXiv251011781P}.
In recent decades, such HVSs have been discovered using data from the Gaia mission, and they have been suggested to have originated from these C- or He-ignited VM systems \citep[][]{2018ApJ...865...15S,2023OJAp....6E..28E,2025MNRAS.541.2231H,2025arXiv251011781P}.
However, the inferred masses of most HVSs are much lower than the donor masses required to trigger dynamical ignition of C or He \citep[$\lesssim 0.4\ M_{\odot}$; see ][]{2025ApJ...982....6S}.

\par

Discussion in Section \ref{subsec:he_vm} suggests that the donor is at least partially disrupted before being ejected as an HVS in both cases of the C- and He-ignited VMs \citep[e.g., ][for a case of the C-ignited VM]{2025arXiv251011781P}. 
This implies that the resulting HVSs could be much less massive than the initial donor masses. 
Thus, this scenario may be consistent with the observed low-mass HVSs. 
Furthermore, the present work showed that the C- and He-ignited VM-driven SNe Ia should commonly form CSM, which well explains the early excesses seen in 03fg/02es-like SNe Ia. 
Hence, we connect the origins of the HVSs to 03fg/02es-like SNe Ia. 
\citet{2025ApJ...982....6S} have estimated that the observed number of HVSs corresponds to $\sim2\ \%$ of the total SNe Ia rate.
This matches the combined event rate of 03fg- and 02es-like objects, $1.3\%$ of all SNe Ia ($\sim0.78\%$ and $\sim0.51\%$, respectively; \citealp{2025A&A...694A..10D}). 
Therefore, the population of 03fg/02es-like objects could account for most, if not all, of the observed HVSs.

\section{Summary}\label{sec:summary}
    In the present study, we have developed an orbital evolution model for the DD binary systems toward the explosion. The model incorporates a wind-driven mass loss induced by the super-Eddington mass transfer.
    Our findings are summarized as follows: \par
    \begin{enumerate}
    \item  The duration of the super-Eddington mass transfer phase ($\gsim10^{-5}\ M_{\odot}\ \mathrm{yr^{-1}}$) is $\sim0.1\text{--}1.0$ years before the coalescence, for a range of the binary parameters. 
        For a lower mass ratio, the duration of the super-Eddington mass transfer phase is longer due to a lower efficiency in the GW-driven mass transfer.\par
    \item  A fraction, $\sim10\%$, of the transferred mass could be ejected, and the accumulated CSM mass is found to be $\sim10^{-2}\ M_{\odot}$ for a range of the binary parameters. 
        We predict the universal CSM density profile; $\rho_{\mathrm{CSM}}\simeq D(r/10^{14}\ \mathrm{cm})^{-3.5}\ (D\simeq 10^{-14}\text{--}10^{-13}\ \mathrm{g\ cm^{-3}})$ when the WDs merge, where the CSM density is higher for a lower mass ratio. 
    \end{enumerate}
        We have then conducted LC synthesis simulations for the SN-CSM interaction, assuming prompt detonation. The results were compared with the observational data.
        Also discussed were carbon emission lines and X-ray emission, which we propose to be a robust tracer of our CSM model.
        Our findings here are summarized as follows:\par
    \begin{enumerate}        
    \item The shock break out radius is $\sim10^{13}\ \mathrm{cm}$ in the context of the VM scenario, i.e., $\sim 0.1$ day after the explosion. The LC as a result of the CSM interaction behaves in a way analogous to the envelope-cooling emission.  
    \item  The early flux excess peaks in a few days after the explosion, both in the $g$-band and in the pseudo-bolometric (3250-8900 {\AA}) luminosities.
        The duration of the CSM interaction is longer for a lower mass ratio, because of the higher CSM density. 
        The peak absolute magnitude of early flux excess in the $g$-band is $\sim-15\text{--} -16$ mag. 
    \item The LCs in the $uvw2$-band also show the early flux excess reaching $\sim-16\text{--}-17$ mag,  which is brighter than that seen in the optical bands. 
        The bright UV excesses help us distinguish between the CSM interaction scenario and other scenarios, e.g., the He-detonation or the companion interaction.
    \item Our models show general agreement with the properties of the early excesses seen in 03fg- and 02es-like objects across optical to UV bands. 
    \end{enumerate}
    Furthermore, we have discussed a case of the He-ignited VM.
    This can be summarized as follows:\par
    \begin{enumerate}
    \item  Our WD-WD binary model predicts that the He-ignited VM does not occur before the onset of tidal disruption of the donor.
    This leads to the CSM formation through the super-Eddington mass transfer and mass ejection, prior to the He-ignition; hence, the He-ignited VM-driven SNe Ia may also exhibit the early excesses.
    Thus, the origin of 03fg/02es-like objects might be a mixture of the C- and He-ignited VMs.
    \item  In both of the C- and He-ignited VMs, the donor is at least partially disrupted before the dynamical ignition. Therefore, even if the donor WDs would survive and ejected from the system as HVSs, their masses should be reduced from the initial masses before the dynamical mass transfer. This may explain the low-mass nature of the observed HVSs. 
    \item We thus connect the 03fg/02es-like SNe Ia and the HVSs through the C- and He-ignited VMs. The number of HVSs discovered by Gaia and the combined rate of 03fg/02es-like SNe Ia match reasonably well. 
    \end{enumerate}

\section{Notation of Physical Constants in this paper}\label{sec:phyconst}
$G$: gravitational constant\par
$m_{\mathrm{e}}$: mass of a electron\par
$m_{\mathrm{n}}$: mass of a nucleon\par
$h$: Planck constant\par
$c$: speed of light\par
$k_{\mathrm{B}}$: Boltzmann constant\par

\appendix

\section{The slope of the CSM density}\label{app:csmslope}
We show that the steep density distribution, $\rho \propto r^{-3.5}$, is a universal feature for the CSM formed by the super-Eddington wind during binary mass transfer, thus can be used as a robust indicator to test our scenario. Within the CSM, the distance from the binary progenitor system at the time of the merger ($r$) is related to the time in the mass-loss history measured backward from the merger ($\widetilde{t}\equiv t_{\rm merge}-t$) through $r=v_{\rm wind}\widetilde{t}$. Therefore, the profile of $\rho\propto r^{-3.5}$ implies an increasing mass-loss rate toward the merger as $\dot{M}_{\rm wind}\propto \rho (r) r^2
\propto r^{-1.5}\propto \widetilde{t}^{-1.5}$ assuming a constant wind velocity. 

Below, we show that this increasing mass-loss rate, $\dot{M}_{\rm wind}\propto \widetilde{t}^{-1.5}$, is generally expected in our scenario. 
We note that most physical quantities, such as $M_{\rm A}$, $M_{\rm D}$, and $a$, change only fractionally during the evolution as shown by our numerical results. 
It is useful to decompose a physical quantity $Q$ into a constant initial value $Q_0$ and a small evolving component $\delta Q$.
For the orbital separation, we define $a_0$ as the value at which the mass transfer rate vanishes. 
Then, the mass transfer rate is determined by $\dot{M}_{\rm D}\propto \left({\cal A}\delta M_{\rm D}+{\cal B}\delta a\right)^3$, which comes from the term $(R_{\rm D}-R_{\rm L})^3$ in Eq.~\eqref{eq:masstransrate}. 
Here $\cal A$ and $\cal B$ are factors that depend only on the initial values.  
In the final phase when the orbital evolution is driven by the mass transfer, 
we have $\delta a\propto \delta M_{\rm D}$ (see Eq.~\ref{eq:orbitalevolution}), and hence $\dot{M}_{\rm D}\propto (\delta M_{\rm D})^3$. 
Then, we have an ordinary differential equation expressed as follows:
\begin{align}
\frac{d(\delta M_{\rm D})}{dt}\propto (\delta M_{\rm D})^3\ .
\end{align}
This gives $\delta M_{\rm D} \propto \widetilde{t}^{-1/2}$. We thus recover the wind mass loss rate as $\dot{M}_{\rm wind}\propto \dot{M}_{\rm D}\propto \widetilde{t}^{-1.5}$.

\section*{acknowledgments}
%-----------------------------------------------------------------------------%
The authors thank Kazuya Iwata, Kohki Uno and Issei Murata for valuable discussion. YI acknowledges financial support from Grant-in-Aid for the Japan Society for the Promotion of Science (JSPS) Fellows (25KJ1472). 
KM acknowledges support from the JSPS KAKENHI grant JP24H01810 and JP24KK0070, and by The Kyoto University Foundation. 
TM acknowledges support from the JSPS KAKENHI grant 24K17088.
\bibliography{sample631}{}
\bibliographystyle{aasjournal}

\end{document}